\documentclass[10pt]{article}
\usepackage{graphicx}
\usepackage[amssymb]{SIunits}

\begin{document}

\title{Controlling the shape of a tapered nanowire: lessons from the Burton-Cabrera-Frank model}

\author{E Bellet-Amalric$^1$, R Andr\'{e}$^2$, C Bougerol$^2$, \\
M den Hertog$^2$, A Jaffal$^{3,4}$, J Cibert$^{2,a}$}

\maketitle


$^1$ Univ Grenoble-Alpes, CEA, IRIG, 38000 Grenoble, France\\
$^2$ Univ Grenoble-Alpes, CNRS, Institut N\'eel, 38000 Grenoble, France\\
$^3$ Univ Lyon, INSA Lyon, CNRS, INL, 69621 Villeurbanne, France\\
$^4$ Univ Lyon, Ecole Centrale de Lyon, CNRS, INL, 69134 Ecully, France\\
$^a$ joel.cibert@neel.cnrs.fr

\vspace{10pt}

\begin{abstract}

The propagation of sidewall steps during the growth of nanowires is calculated in the frame of the Burton-Cabrera-Frank model. The stable shape of the nanowire comprises a cylinder section on top of a cone section: their characteristics are obtained as a function of the radius of the catalyst-nanowire area, the desorption-limited diffusion length of adatoms on the terraces, and the sticking of adatoms at step edges. The comparison with experimental data allows us to evaluate these last two parameters for InP and ZnTe nanowires; it reveals a different behavior for the two materials, related to a difference by an order of magnitude of the desorption-limited diffusion length.

\end{abstract}

\vspace{2pc} \noindent{\it Keywords}: nanowires, steps,
semiconductors, molecular beam epitaxy, electron microscopy, Burton-Cabrera-Frank model

%
%

\section{Introduction}

The interest of a tapered waveguide for directing the light emitted into free space by a single photon emitter such as a quantum dot (QD) has been first demonstrated thanks to structures realized by a top-down approach, by etching a GaAs wafer containing multiple InAs quantum dots \cite{Claudon2010}. Quite rapidly however, the possibility to obtain a suitable waveguide in a bottom-up process during the growth of a tapered nanowire (NW) was demonstrated by positioning a single (In,As)P QD in an InP NW \cite{Reimer2012, Dalacu2012, Haffouz2018}. In addition to the exact positioning of the QD on the NW axis, this allows a further manipulation of the NW-QD ensemble in order to insert it into a photonic circuit \cite{Zadeh2016, Mnaymneh2019} in which different active components can be inserted. The efficiency of the coupling to the air or to the waveguide depends on the cone angle, with typical values of the order of a few degrees \cite{Mnaymneh2019, Jaffal2019}, and typical length a few micrometers.

Growing a NW with a regular cone shape may seem a simple task, as it should be enough to adjust and control a constant ratio between the radial growth rate $dR/dt$ and the axial growth rate $dL/dt$. This however involves complex entangled mechanisms, since the vapor-liquid-solid (VLS) or vapor-solid-solid (VSS) mechanism for the axial growth, and the vapor-solid mechanism for the radial growth, are fed and linked by diffusion processes. As a result, more complex shapes are experimentally observed, as for instance pencil-shaped NWs \cite{Tchernycheva2007}, or other shapes \cite{Dubrovskii2008, Dubrovskii2013}. Most of the theoretical studies have been devoted to quantifying the axial growth rate, with the accent put first on the diffusion processes to the droplet or nanoparticle acting as a catalyst \cite{Dubrovskii2006}; the role of the nucleation and propagation of monoatomic steps at the liquid-solid or solid-solid interface is now well recognized and studied \cite{Hofmann2008, Wen2010, Harmand2018}. Sophisticated models have been applied to one of the simplest systems: atomistic simulations developed for the metal-covalent system \cite{Dongare2009} point to the role of facets and their variation during the growth of Si NWs \cite{Wang2013}. The line tension at the droplet-NW interface \cite{Checco2003, Weijs2011} determines the contact angle and hence the shape of the droplet and the size of the contact area: as a result, a modification of the NW radius may be expected during the early step of growth \cite{Schmidt2005}. Turning to the radial growth, descriptions generally focus onto the nucleation on the sidewalls \cite{Dubrovskii2008, Chen2001, Plante2009}, although step flow is often invoked. Step dynamics has been explicitly taken into
account more recently \cite{Filimonov2015, Filimonov2016}, and examples of parameters are given which lead to a cone shape, a pencil shape, or a series of step bunches.

Here we discuss the role of step dynamics on the stabilization of a
cone-shaped, tapered NW, during its growth; we examine the consequences for the actual shape of the NW, which exhibits a cylinder-shaped tip section. Analytical results are presented on the range of values which can be attained for the cone angle and the cylinder length, as well as the axial and radial growth rates. These results are applied to two different systems, InP and ZnTe, for which experimental results are available on the growth of NWs in the VLS and VSS modes respectively.

\section{Experimental aspects}

InP NWs grown by metal organic molecular beam epitaxy (MOMBE) display a cone shape close to the substrate, with a cylinder-shaped section between the cone and the droplet  \cite{Greenberg2014, Kelrich2015}. The length of the cylinder depends on the NW radius \cite{Greenberg2014} and has been considered to be a measure of the so-called migration length. In contrast, short InP NWs grown by solid-source molecular beam epitaxy (MBE) feature a pencil shape \cite{Jaffal2019}; however longer NWs, or NWs with a shell grown at lower temperature, display a regular cone shape on top of a cylinder section \cite{Jaffal2019, Harmand2013}. Large values of the half angle $\theta$ of the cone, with $\tan \theta$ up to 0.16, are obtained by decreasing the growth temperature, which decreases $dL/dt$ while keeping $dR/dt$ almost constant \cite{Jaffal2019, Harmand2013}.

\begin{figure}
\centering
\includegraphics [scale=1.0]{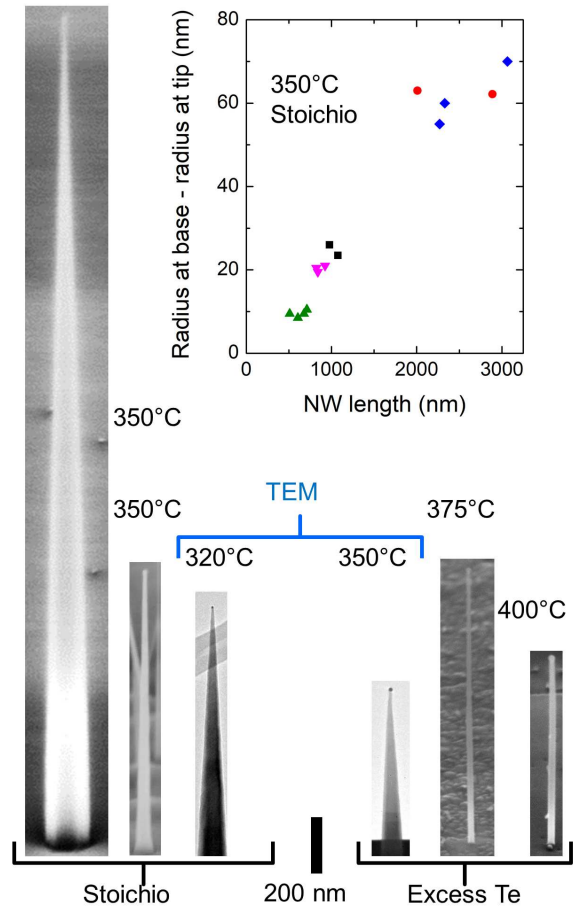}
\caption{ZnTe NWs grown under different conditions of flux ratio (stoichiometric from a ZnTe effusion cell, or adding a Te flux) and temperature (from $320^\circ $C to $400^\circ $C), as indicated. The values of flux and growth times are different for each NW (resulting for instance in a factor of about 3 in the length of the two NWs shown on the left). The NWs are observed by SEM (tilt $65^\circ$), except for the two central ones observed by TEM. The scale bar (200 nm) is common to all images. Insert: increase of the radius as a function of length, for NWs grown at $350^\circ $C under stoichiometric ZnTe flux, with different values of the flux and growth time (several NWs for each sample, with a specific symbol for a given sample); the slope gives the aspect ratio.} \label{fig1}
\end{figure}

Figure~\ref{fig1} introduces our second example, that of ZnTe NWs, grown by MBE, and observed by scanning electron microscopy (SEM) with a 65$^\circ$ tilt, or transmission electron microscopy (TEM): growth and observation conditions are described in Ref.~\cite{Rueda2014} and \cite{Rueda2016}. The NWs shown in Fig.~\ref{fig1} were chosen as having  a gold nanoparticle radius $\sim 5$~nm (except the NW grown at 400$^\circ$C since the nanoparticle size is larger at this temperature); all have the zinc-blende structure. Taking the growth at 350$^\circ$C with stoichiometric flux as an example, the two images and the plot of the radius-length dependance illustrate the reproducibility of the aspect ratio for different growth times or flux values. Such a stable character of the shape during the growth was already noticed even for more complex shapes of III-V NWs \cite{Chen2006}. Figure~\ref{fig1} illustrates the possibility to adjust the cone angle by the control of the growth conditions: it will be shown below that the increase of angle value is related to a concomitant increase of the radial growth rate and decrease of the axial growth rate.

\section{Model}

A regular cone shape suggests the presence of a regular step array on the
sidewalls, and the dynamics of these steps must be taken into account. A
simplified view involves well-formed (110) facets or a circular
cross section. Although the actual picture certainly involves a more complex shape and some disorder, we assume that the step height is that of (110) facets,
$a_\bot=a_0/2\sqrt{2}$ where $a_0$ is the zinc-blende lattice
parameter. We note $a_\|$  the interplane distance perpendicular to the
step ledge in the terrace plane (\emph{i.e.}, along the NW axis), $a_\|=a_0/\sqrt{3}$.

Our main assumption is the availability of steps nucleated at the basis of the NW, or in a section below the expected cone-shaped section.

The Burton-Cabrera-Frank model (BCF) \cite{BCF} is commonly used to
describe growth by MBE \cite{PimpinelliVillain}.
The flux of atoms creates adatoms which diffuse on terraces (as
described by the diffusion coefficient $D$, associated to a hopping
probability $p_{hop}$ of an adatom to a neighboring site), until they desorb
(with a probability of desorption $p_{des}$ and a lifetime $\tau=\frac{1}{p_{des}}$) or are incorporated into the crystal when they encounter a proper
nucleation center. Steps form such nucleation centers, and the
simplest formulation of growth by step flow assumes a
one-dimensional array of steps, separated by terraces. While these
concepts have been widely used to describe the growth of
two-dimensional layers, their application to the radial growth and
the dynamics of steps on NW sidewalls, towards a better
understanding of the shape of these NWs, is quite recent
\cite{Filimonov2015, Filimonov2016}. In this series of papers, the
sticking of adatoms at steps was assumed to be large. A final
numerical calculation was performed for chosen values of the characteristic
parameters describing the dynamics of steps: it allowed the authors to demonstrate the
possibility of a cylinder shape with radial growth, of a "pencil
shape" due to step bunching, and of tapering \cite{Filimonov2015, Filimonov2016}. The nucleation of
steps was shown to take place at or near the basis of the NW. In the present study, we introduce the possibility of a low probability of
incorporation at the steps \cite{Hata1991}, $p_{inc}$, which may be much lower
than the hopping probability $p_{hop}$ at the origin of diffusion on the
terraces. We also explicitly take into account the desorption of
adatoms and the sublimation of the material, two effects which are
currently observed during the MBE growth of II-VI semiconductors,
and were taken into account when using the BCF formalism to describe
the growth of a CdTe layer \cite{Pimpinelli1998, Peyla1998}.

\begin{figure}
\centering
\includegraphics [scale=1.0]{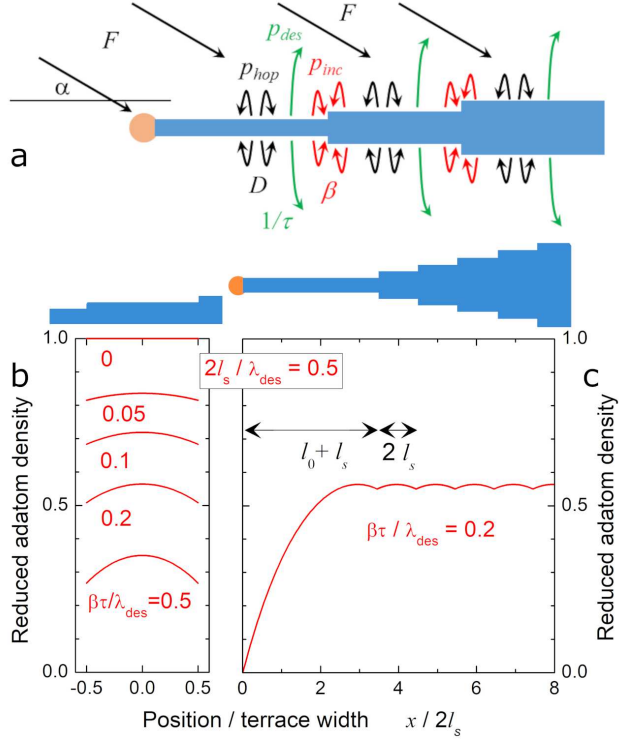}
\caption{(a) Scheme of the BCF model with material parameters; the flux reaches the terraces, at an angle $\alpha$ with respect to the NW axis, and also contributes directly to the concentration in the nanoparticle/droplet; (b) and (c) Reduced adatom density $\frac{n(x)-n_{eq}}{F_\alpha \tau-n_{eq}}$ as a function of the position
according to the BCF model applied (b) to an array of regular steps, and (c) to the step array and NW tip, in both cases with values of $\frac{\beta\tau}{\lambda_{des}}$ and $\frac{2l_s}{\lambda_{des}}$ as indicated.} \label{fig2}
\end{figure}

The material parameters relevant for the BCF model are schematized in Fig.~\ref{fig2}~(a). A list of characteristic lengths is given in Table~\ref{Table}. Steps of the calculation are detailed in the appendix.

The sidewalls of the NW receive an average flux $F_\alpha=F \sin
\alpha / \pi$, where $F$ is the flux along the cell to NW direction,
and $\alpha$ is the angle between this direction and the NW axis. Note
that the calibration of the flux is obtained from the growth rate on
a substrate, which gives a measure of $F \cos \alpha$. It is quite
convenient to express these fluxes in nm per unit time,
corresponding to the growth rate of a thick layer of the material if every atom sticks: that means that $F$ is the flux in atoms per unit area and unit time, multiplied by the volume $\Omega$ of the unit cell.

On a wide terrace and in the absence of nucleation, the incident flux is balanced by
desorption, so that the adatom density (actually, the adatom density
multiplied by the volume $\Omega$ of the unit cell) is $n=F_\alpha
\tau$, where $p_{des}=1/\tau$ is the probability of adatom
desorption.

In its simplest formulation, growth by step flow assumes a
one-dimensional array of steps, separated by terraces of length
$2l_s$ (Fig.~\ref{fig2}a and b). The adatom density at position $x$ along a
terrace, $n(x)$, obeys the diffusion equation characterized by $D$
and $\tau$: the uniform solution $n=F_\alpha
\tau$ is modified by adding a
combination of two exponential or two hyperbolic functions, for
instance $n(x)=F_\alpha \tau-A \cosh(x/\lambda_{des})-B
\sinh(x/\lambda_{des})$, where $\lambda_{des}$ is the diffusion
length limited by the desorption of the adatoms,
$\lambda_{des}=\sqrt{D\tau}$. The two coefficients $A$ and $B$ are
determined by the boundary conditions at the steps: if, for the sake
of simplicity, we assume no Ehrlich-Schw\"{o}bel effect \cite{PimpinelliVillain}, the
symmetry imposes $B=0$ if the position is measured from the center
of the terrace. The parameter $A$ is determined by the fact that the current related to the gradient of adatoms
balances the incorporation of adatoms on both sides at the step:
$\mp D \nabla n (\pm l_s) = \beta [n(\pm l_s)-n_{eq}]$, where the parameter $\beta$ sets the efficiency of trapping at the step \cite{Filimonov2015, Filimonov2014}, and
$n_{eq}$ is determined by the sublimation rate $V_{sub}$ of the material,
$n_{eq}=V_{sub} \tau$. The sublimation process involves the release of an atom already incorporated, generally at a step edge, adding thus an adatom to the neighboring terrace \cite{Pimpinelli1998}, which then diffuses and may be desorbed or re-incorporated.

\begin{table} [h]
\begin{center}
\begin{tabular}{|l|l|}
  \hline
  \textbf{symbol} & \textbf{parameter}  \\
  \hline
  $2l_s$ & terrace length in the cone\\
  \hline
  $l_0+l_s$ & terrace length at tip\\
  \hline
  $\lambda_{des}$ & diffusion length on infinite terrace  \\
  \hline
  $\lambda_{eff}$ & effective diff. length for axial growth rate  \\
  \hline
\end{tabular}
\end{center}
\caption{Characteristic lengths used in the text\label{Table}}
\end{table}

A straightforward calculation leads to

\begin{equation}\label{Eq1}
\frac{n(x)-n_{eq}}{F_\alpha \tau-n_{eq}}=
1-\frac{\frac{\beta\tau}{\lambda_{des}}}{\frac{\beta\tau}{\lambda_{des}}+\tanh(\frac{l_s}{\lambda_{des}})}\frac{\cosh(\frac{x}{\lambda_{des}})}{\cosh(\frac{l_s}{\lambda_{des}})}.
\end{equation}

The incorporation of adatoms at a step induces the propagation of
the step, with a velocity $V_{step}$, and the growth of the layer,
with a growth rate $\frac{dh}{dt}=\frac{a_\bot}{2l_s}V_{step}$. In
the present case of a NW, this is actually the radial growth rate
$\frac{dR}{dt}$. From Eq.\ref{Eq1}, we obtain
\begin{eqnarray}\label{Eq2}
\frac{dR}{dt}=(F_\alpha-V_{sub}) \frac{\lambda_{des}}{l_s}\frac{\beta\tau ~\tanh(\frac{l_s}{\lambda_{des}})}{\beta\tau+\lambda_{des}\tanh(\frac{l_s}{\lambda_{des}})},\nonumber\\
V_{step}=(F_\alpha-V_{sub}) \frac{2\lambda_{des}}{a_\bot}\frac{
\beta\tau~
\tanh(\frac{l_s}{\lambda_{des}})}{\beta\tau+\lambda_{des}\tanh(\frac{l_s}{\lambda_{des}})}.
\end{eqnarray}

The sublimation rate of the semiconductor material (represented by $V_{sub}$ and $n_{eq}$ in the previous equations), must be taken into account for the growth of CdTe in the form of 2D layers \cite{Pimpinelli1998} and NWs \cite{Orru2018}: it results in a decrease of the growth rate when the temperature is increased. We find no  such evidence that it plays a role for ZnTe and InP NWs grown under the present conditions and it will be omitted in the following.

The adatom density on a single terrace separated by two steps is
plotted in Fig.~\ref{fig2}~(b) for different values of the sticking
probability. At strong sticking probability, the boundary condition
is simply $n(l_s)=n_{eq}$; a decrease in the sticking efficiency is
partially compensated by an increase of $n(l_s)$, leading however to
a decrease of the adatom current and a decrease of the step
velocity. The step velocity depends also on the terrace length
$2l_s$. The maximum step velocity, obtained for a low density of
steps and a strong sticking, is $V_{step}=(F_\alpha-V_{sub})
\frac{2\lambda_{des}}{a_\bot}$.

This propagation of the step on the sidewalls of a cone-shaped NW has to be compared to the (axial) growth rate of the NW. The catalyst (nanoparticle or droplet) present at the tip of the NW is often considered as a perfect trap, implying
$n(0)=n_{eq}$ at the NW tip.

In the absence of steps on the
sidewalls in the vicinity of the NW tip, the adatom density is
described by a simple exponential function of $x/\lambda_{des}$; the
NW growth rate is obtained by writing that the adatom current
feeds a disk of radius $R$ at the tip of the NW, hence
$\frac{dL}{dt}=(F_\alpha-V_{sub}) \frac{2\lambda_{des}}{R}$. This is
much smaller than the maximum step velocity
$V_{step}=(F_\alpha-V_{sub}) \frac{2\lambda_{des}}{a_\bot}$ given in
the previous paragraph. Indeed, this is one of the configurations
considered by Filimonov and Hervieu \cite{Filimonov2015,
Filimonov2016}: if both the nanoparticle and the steps behave as
deep traps, since the area $\pi R^2$ of the disk to be filled at the tip is
much larger than the area $2 \pi R a_\perp$ of the ring to be filled at a step, each
step nucleated at the base of the NW rapidly reaches the tip
and the NW acquires a cylinder shape with increasing radius.

If the sticking at steps is small enough, a terrace length may be
found so that the step velocity in the section with a regular cone shape, matches the NW growth rate, so that a stable configuration can be contemplated. It remains to be decided whether the intermediate zone between the regular cone section and the tip can host steps. In this section, the adatom density decreases from its value in the cone section, with $n$ given by Eq.~\ref{Eq1}, and the
NW tip with $n(0)=n_{eq}$: with an intermediate value of $n$,
the velocity of these hypothetical steps is lower than the velocity
of the steps in the regular array, which will thus catch them up; it is also lower than the NW growth rate, so that the NW tip will escape. The
resulting configuration is thus the array of equidistant steps, with
a longer terrace at the tip, of length $l_s+l_0>2l_s$, Fig.~2~(c). On this asymmetric terrace (with different boundary conditions on both sides), a general procedure is to write the adatom density as a linear combination of a cosh and a sinh functions. Actually, it is as general to write it using a single cosh (or sinh) function, with the extremum not centered on the terrace. In the present case, the boundary conditions are easily satisfied by using the same function, Eq.~\ref{Eq1}, with the origin at $l_s$ from the leading step, and the second part extending to the tip of length
$l_0$ such that $n(0)=n_{eq}$; hence

\begin{equation}\label{Eq3}
\frac{\frac{\beta\tau}{\lambda_{des}}}{\frac{\beta\tau}{\lambda_{des}}+\tanh(\frac{l_s}{\lambda_{des}})}\frac{\cosh(\frac{l_0}{\lambda_{des}})}{\cosh(\frac{l_s}{\lambda_{des}})}=1.
\end{equation}

The gradient of adatom density at the NW
tip allows us to calculate the NW growth rate as
\begin{eqnarray}\label{Eq4}
\frac{dL}{dt}=(F_\alpha-V_{sub}) \frac{2\lambda_{des} \tanh(\frac{l_0}{\lambda_{des}})}{R}+F~\kappa,\nonumber\\
=(F_\alpha-V_{sub}) \frac{2\lambda_{eff}}{R}.
\end{eqnarray}

The last part defines an effective diffusion length, $\lambda_{eff}$, which gives to the usual expression of the growth rate as a function of the "diffusion length", $\frac{dL}{dt}=F \frac{2\lambda}{R}$, a precise meaning. The term $F~\kappa$ accounts for the contribution of the direct flux to the catalyst. The value of $\kappa$ depends on the shape of the droplet or nanoparticle \cite{Glas2010} and will be discussed in Section \ref{compar} for InP and ZnTe NWs. For a NW of radius $R$ with material parameters described by
$\lambda_{des}$ and $\beta\tau/\lambda_{des}$, the parameters $l_s$,
$l_0$ and $\lambda_{eff}$ characterizing a stable configuration are
obtained by equating the growth rate (Eq.~\ref{Eq4}) and the step
velocity (Eq.~\ref{Eq2}). Straightforward calculations lead to analytical expressions.

For an analysis of experimental data, as done in Section \ref{compar}, it is interesting to eliminate $\beta \tau /\lambda_{des}$ from $V_{step}$ (Eq.~\ref{Eq2}), using the boundary condition (Eq.~\ref{Eq4}), so that the relation $dL/dt=V_{step}$ takes the form

\begin{eqnarray}\label{Eq5}
\frac{\sin \alpha}{\pi} \frac{2\lambda_{des}}{R}[\tanh(\frac{l_0}{\lambda_{des}})-\frac{R}{a_\perp} \sinh(\frac{l_s}{\lambda_{des}})\frac{1}{\cosh(\frac{l_0}{\lambda_{des}})}]\nonumber\\
+\kappa =0.
\end{eqnarray}

This equation combines two measurable parameters ($l_s$, \emph{i.e.}, the cone angle $\theta$ with $\tan \theta=\frac{a_\bot}{2 l_s}$, and $l_0$, the cylinder length) and contains a single (\emph{a priori} unknown) material parameter $\lambda_{des}$. It is easily cast into a second-degree equation for $\tanh(\frac{l_0}{\lambda_{des}})$ which will be used in Section \ref{compar}.

For a discussion of the mechanisms, as done in Section \ref{discussion}, it is useful to re-arrange the previous equations in order to write the measurable parameters as a function of the material parameters. If the contribution of the direct flux to the catalyst is negligible ($\kappa=0$), a straightforward calculation to eliminate $l_0$ leads to quite simple expressions for $l_s$, and in turn for $l_0$:

\begin{eqnarray}\label{Eq6}
\tanh(\frac{l_s}{\lambda_{des}})=\frac{2\frac{\beta\tau}{\lambda_{des}}}{\left[(\frac{R}{a_\bot})^2-1\right](\frac{\beta\tau}{\lambda_{des}})^2-1}\nonumber\\
\tanh(\frac{l_0}{\lambda_{des}})=\frac{R}{a_\bot}\frac{2\frac{\beta\tau}{\lambda_{des}}}{\left[(\frac{R}{a_\bot})^2-1\right](\frac{\beta\tau}{\lambda_{des}})^2+1}
\end{eqnarray}

One may note that the material parameters appearing in Eq.\ref{Eq6} are $\lambda_{des}$ and $\frac{\beta\tau}{\lambda_{des}}$; in other words, the relevant material parameters are $\lambda_{des}$ and $\frac{\beta\tau}{\lambda_{des}^2}$, and it is useful to remember that the ratio of these quantities to the hopping distance are respectively $(p_{des}/p_{hop})^{1/2}$ and $(p_{inc}/p_{hop})$.

\section{Discussion} \label{discussion}

In this section we limit ourself to the case with $\kappa=0$ (negligible direct flux to the catalyst), and use Eq.~\ref{Eq6} relating the experimental parameters to the material parameters $\lambda_{des}$ and $\beta\tau/\lambda_{des}$.

Apart from the terrace length in the cone section, $2 l_s$, we thus have three characteristic lengths which have to be clearly identified (Table~\ref{Table}): the diffusion length $\lambda_{des}$ limited by desorption on a long terrace, the length $l_0$ describing the excess length of the cylinder section (its length is $l_0+l_s$), and the effective diffusion length $\lambda_{eff}$ (defined in Eq.~\ref{Eq4}) describing the contribution to the axial growth rate.

\begin{figure}
\centering
\includegraphics [scale=1]{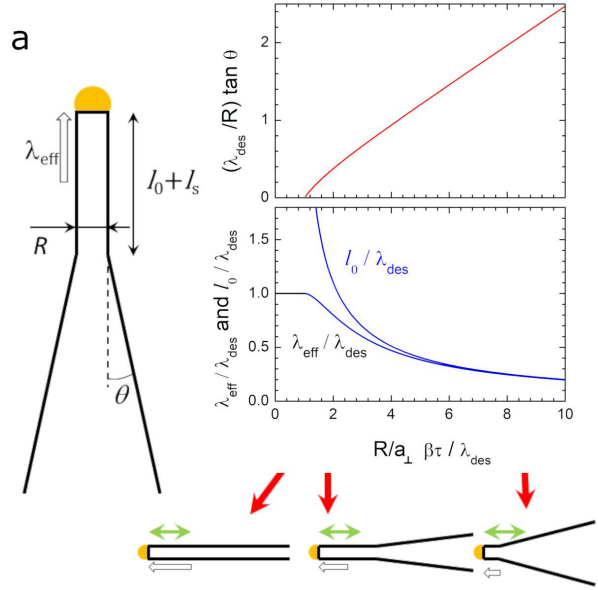}
\caption{(a) Measurable parameters describing the NW shape;
(b) Plot of the cone half-angle as a function of
the sticking, in reduced units (see text), assuming a fixed value of the
desorption-limited diffusion length $\lambda_{des}$; (c)
diffusion length to the tip and length of the tip terrace. The scheme at the bottom represents the three typical cases described in the text: cylinder-shape, cone
with a small angle, cone with a large angle. The green arrows schematize $\lambda_{des}$ and the white arrows $\lambda_{eff}$.} \label{fig3}
\end{figure}

Figure \ref{fig3} shows a plot of
$\frac{l_0}{\lambda_{des}}$, $\frac{\lambda_{eff}}{\lambda_{des}}$
and $\frac{\lambda_{des}}{R}~\tan \theta$, as a function of
$\frac{R}{a_\bot}~\frac{\beta\tau}{\lambda_{des}}$. The plot is for $R$=5~nm, $a_\bot=0.22$~nm and $\lambda_{des}=100$~nm but the reduced plot is quite valid also for other values.

\begin{itemize}
  \item In the absence of sticking at steps, the NWs feature a
cylinder shape ($\theta=0$) and the diffusion to the nanoparticle is
determined by adatom desorption (\emph{i.e.}, $\lambda_{eff}=\lambda_{des}$).
  \item This remains true for small values of the sticking, up to a threshold
$\sim\frac{a_\bot}{R}$ (left scheme at the bottom of Fig.~\ref{fig3}). Then
$\tanh(\frac{l_s}{\lambda_{des}})\approx\tanh(\frac{l_0}{\lambda_{des}})\approx1$,
and for $\frac{\beta\tau}{\lambda_{des}}<\frac{a_\bot}{R}$ the step
velocity, $V_{step}\approx(F_\alpha-V_{sub})
\frac{2\beta\tau}{a_\bot}$ is smaller than the growth rate
$\frac{dL}{dt}\approx(F_\alpha-V_{sub}) \frac{2\lambda_{des}}{R}$.
It implies that steps nucleated at the base of the NW remain
stuck there, while a cylinder-shaped section emerges and gains in length, forming an essentially non-tapered NW. We tentatively
ascribe ZnTe NWs with the wurtzite structures \cite{Rueda2014}, as well as zinc-blende ZnTe
NWs grown at high temperature (see next section) to this regime.
  \item Above this threshold, radial growth takes place. \begin{itemize}
          \item Well above the
threshold (right scheme at the bottom of Fig.~\ref{fig3}), the shape is governed by sticking at the step edges:
$l_0$, $l_s$ and $\lambda_{eff}$ are all much smaller than
$\lambda_{des}$. Then Eq.~\ref{Eq4} and \ref{Eq6} can be simplified
into
$l_0\approx\lambda_{eff}\approx\frac{2a_\bot}{R}~\frac{D}{\beta}$
and $\tan \theta\approx\frac{R^2}{4a_\bot}\frac{\beta}{D}$, with a
simple relationship between the cone angle, the radius and the
length of the tip terrace. The examples of InP NWs and most of the ZnTe NWs examples given in the next section pertain to this range. Note that $\frac{\beta}{D}=\frac{1}{a_{\|}}\frac{p_{inc}}{p_{hop}}$.
          \item When approaching the threshold from above (central scheme at the bottom of Fig.~\ref{fig3}), $l_0$ diverges while
$\lambda_{eff}$ is limited by desorption and approaches
$\lambda_{des}$. One example of ZnTe NWs is close to this limit.
        \end{itemize}
  \end{itemize}

\section{Comparison with experimental data} \label{compar}

\begin{figure}
\centering
\includegraphics [scale=1.0]{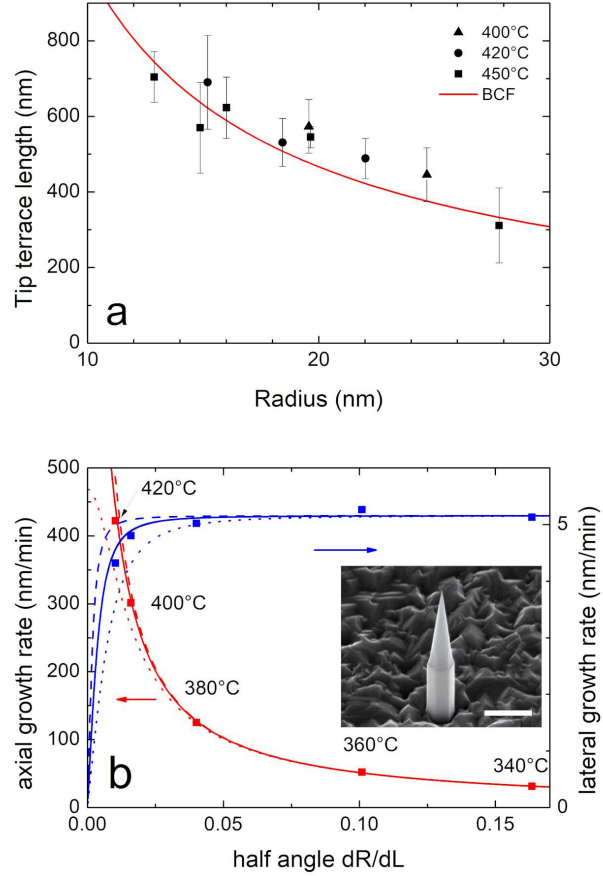}
\caption{InP NWs. (a) MOMBE growth: symbols show the "migration length" measured in Ref.~\cite{Greenberg2014}; the line shows $l_0$ calculated according to
Eq.~\ref{Eq6}, with $\lambda_{des}$=1000~nm and $\beta \tau /\lambda_{des}$=0.05; (b) solid-source MBE growth: symbols show the axial and radial growth rates measured on the cone section of core-shell NWs, as a function of tapering ($dR/dL=\tan \theta$); solid lines are the fit (see text) for $\lambda_{des}$=1000~nm, dotted lines for $\lambda_{des}$=500~nm, dashed lines for $\lambda_{des}$=2000~nm; the inset shows a SEM image (tilt 45$^{\circ}$, scale bar 1~$\mu$m) of an InP core-shell NW with the shell grown at 360$^{\circ}$C.} \label{fig4}
\end{figure}

The presence of a long cylinder-shaped segment below
the tip of tapered InP NWs grown by MOMBE was studied experimentally in Ref.~\cite{Greenberg2014}. Other examples of this shape can be found in Ref.~\cite{Kelrich2015}. The length of
this segment (called "migration length") was measured on TEM images of NWs collected on a carbon grid \cite{Greenberg2014}: it significantly depends on the diameter of the NW,
see the experimental data in Fig.~\ref{fig4}~(a). This diameter-dependence of the "migration length" was kept as an adjustable parameter in a subsequent model of
the NW shape \cite{Dubrovskii2015} which takes into account
nucleation effects at the base of the NW. The solid line in
Fig.~\ref{fig4}~(a) plots $l_0$ calculated using Eq.~\ref{Eq6}, with $\lambda_{des}$=1000nm and $\beta \tau /\lambda_{des}$=0.05. Actually, a good fit is obtained assuming a value of the desorption-limited diffusion length $\lambda_{des}$ in the micrometer range, as expected \cite{Greenberg2014}, provided the relevant parameter $\beta \tau
/\lambda_{des}^2$ (\emph{i.e.}, the ratio of incorporation probability to hopping probability, $p_{inc}/p_{hop}$, as discussed above) is conserved. In the present plot, this parameter is of course independent of the NW diameter. Unfortunately, the cone of these NWs is not regular\cite{Greenberg2014, Kelrich2015}: this is probably due to the fact that this section is rather short, so that nucleation effects play a significant role, as discussed specifically in Ref.~\cite{Dubrovskii2015}; this aspect is not included in the present study.

A more regular shape was reported for NWs grown by solid-source MBE \cite{Jaffal2019, Harmand2013}. In this case, short NWs feature a pencil-like shape, which is not the focus of the present study. Longer NWs, or NWs involving a final step at low temperature in order to form a shell, exhibit a cone shape and the present model can be used. NWs with the shell grown at different temperatures were observed by SEM (see the inset of Fig.~\ref{fig4}b, and Ref~\cite{Jaffal2019} for other examples), allowing us to deduce the change of NW radius and length due to the shell-growth step (hence only the cone section is taken into account). Fig.~\ref{fig4}~(b) shows the dependance of the axial and radial growth rates, $dL/dt$ and $dR/dt$, on the tangent of the half angle \cite{Jaffal2019}. These two plots are not independent since $\tan \theta = dR/dL$, but it is instructive to plot them separately. A good fit is obtained using Eq.~\ref{Eq4} and \ref{Eq5} to calculate $dL/dt$, with only two adjustable parameters: the value of the flux, which determines the vertical scale, and the diffusion length due to desorption, $\lambda_{des}$.  The experimental behavior of $dL/dt$ and $dR/dt$ is reproduced using a constant value $\lambda_{des}$=1000~nm, with an increase of the sticking when decreasing the temperature (not shown), starting from a value which coincides with that of Fig.~\ref{fig4}~(a) to within a factor 2. The evolution of the tapering angle is induced by the change of the axial growth rate, with a change of radial growth which is very small and limited to the high temperature range. Increasing or decreasing the value of $\lambda_{des}$ rapidly modifies the radial growth rate at high temperature, while the low temperature range is practically unchanged (the radial growth is at saturation). Figure~\ref{fig4}~(b) also reveals the almost negligible contribution from the direct flux. This is a consequence of the large value of $\lambda_{des}$, for a part, but also of the geometrical configuration: the flux from the cell makes a large angle, $\alpha \sim 45^\circ $, with the NW axis, and the droplet is quasi-flat \cite{Jaffal2019}. In these conditions \cite{Glas2010}, $F~\kappa \simeq F \cos \alpha = F/\sqrt{2}$, and $F \sin \alpha = F/\sqrt{2}$. In Fig.~\ref{fig4}~(b), the curves calculated with $\kappa=0$ and $\kappa\neq0$ are undistinguishable.

\begin{figure}
\centering
\includegraphics [scale=1]{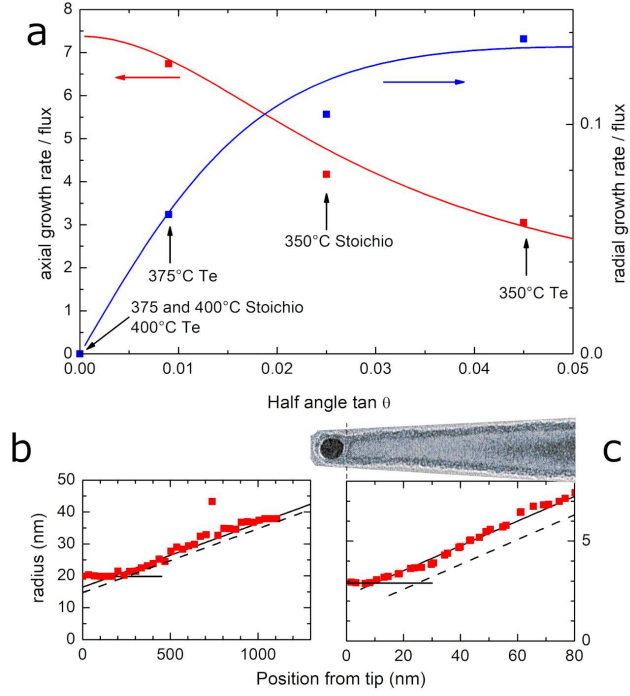}
\caption{ZnTe NWs. (a) axial and radial growth rates, as a function of tapering (tangent of the half angle), and fit (see text). (b) Radius as a function of the distance to the tip, measured on Fig.~1b of Ref. \cite{Wojnar2012}. (c) TEM image and plot, at the same horizontal scale, of the radius as a function of the distance to the tip, sample grown at $320^\circ$C under stoichiometric flux.}\label{fig5}
\end{figure}

Turning now to ZnTe, the detailed study of the axial growth rate \cite{Rueda2016, Orru2018} allowed us to estimate the effective diffusion length $\lambda_{eff}$. When measured at different temperatures on the same, single NW \cite{Orru2018}, $\lambda_{eff}$ is characterized by a monotonous increase with temperature, but also a minimum value at low temperature which we attribute to the direct flux. In a previous study \cite{Rueda2016}, we used the radial growth in order to determine the nucleation delay and disentangle the axial growth of the NW, for NWs grown under stoichiometric ZnTe flux and with an excess of Te. We also confirmed the strong contribution from the flux to the nanoparticle, with a quasi-spherical profile and a contact area much smaller than the NW section: we estimate $\kappa \simeq 2.5$, while $\sin \alpha = 0.43$. In addition, we may expect a significant desorption of the adatoms, as observed in the growth of thick layers \cite{Pimpinelli1998}. It is thus mandatory to take into account the direct flux. Figure~\ref{fig5}~(a) uses the axial growth rate, taken from Ref~\cite{Rueda2016}, and the angle taken from the present Fig.~\ref{fig1}: it shows that the angle is increased (by decreasing the temperature or adding a Te flux) by a combination of the increase of the radial growth and the decrease of the axial growth. A good fit is obtained by keeping $\lambda_{des}$ constant at 90 nm: this implies a change of the sticking efficiency (\emph{i.e.}, of $p_{inc}/p_{hop}$). The role of temperature is further confirmed by the fact that NWs grown at higher temperature (at $375^\circ$C or $400^\circ$C under stoichiometric ZnTe, at $400^\circ$C under Te excess) exhibit no radial growth, see Fig.~\ref{fig1}, as expected for conditions below the threshold discussed in Section~\ref{discussion}.

A cylinder-shaped tip section is visible in Fig~1~(b) of
Ref~\cite{Wojnar2012}: the NW was grown directly on the
GaAs substrate, the growth mode is VLS and the radius at tip is quite large ($R$=20 nm). An analysis of the image, Fig.~\ref{fig5}~(b), shows that the tapered section starts approximately 230 nm below the tip, in agreement with the present model for the measured half-angle value, $\tan\theta\approx0.022$. The tip section is expected to be much smaller on NWs with a smaller tip radius (as we obtain on a ZnTe buffer layer) and a stronger tapering. In this case, SEM images are not sufficient as TEM reveals the presence of a 2 to 3~nm thick amorphous layer, which masks the crystalline core (see Fig.~\ref{fig5}~(c)). When the core can be identified in spite of the high sensitivity of these thin NWs to the electron beam, a tip section can be observed (Fig.~\ref{fig5}~(c)) with a length in good agreement with the value calculated for the measured values of the radius and cone angle.

The influence of the direct flux to the nanoparticle is quite sizable: in the absence of this direct flux, the cylinder section would be longer, see the dashed line in Fig.~\ref{fig5}~(b) and (c): in this case ($\kappa=0$ in Eq.\ref{Eq4}), the axial growth rate is fully ensured by the sidewall flux through the tip terrace. For a given value of $dL/dt$, adding a contribution from the direct flux to the nanoparticle thus reduces $l_0$. Actually the presence of a direct flux sets an upper limit to the tapering angle which can be reached by increasing the sticking at steps. At diverging $\frac{\beta\tau}{\lambda_{des}}$, and for $\frac{l_s}{\lambda_{des}}
\ll1$, Eq.~\ref{Eq2} is simplified into $\frac{dR}{dt}=F_\alpha$, which is obviously the upper limit for the radial growth rate; with a lower limit $\frac{dL}{dt}=\kappa~F$ for the axial growth rate, an upper limit to the tapering angle is obtained at $\tan \theta=\frac{\sin \alpha}{\pi~\kappa}$.

Our model assumes that nucleation takes place at the base
of the NW, or at least over the bottom section. It appears to be the case for the ZnTe NWs grown at low temperature, which offer a cone shape over their whole length. When a core-shell NW, made of ZnTe or InP, is realized by starting with a NW grown at high temperature, and finalizing the structure with a second growth step at low temperature, the steps are nucleated on the sidewalls of the initial NW. An interesting case is that of InP NWs grown in the high temperature range \cite{Jaffal2019, Harmand2013}: they grow initially with a pencil-like shape, with a rather low radial growth rate. Then, the radial rate increases and they acquire a cone shape. We tentatively ascribe the initial pencil-like shape to a lack of nucleation of steps at the base of the NW, perhaps related to the adatom flux from the sidewalls to the substrate, and not from the substrate to the NW \cite{Harmand2010}. It is only when the sidewalls are long enough that steps nucleated over the whole length \cite{Dubrovskii2008, Chen2001, Plante2009} allow the formation of the cone section. The present analysis also suggests that the difference between the InP NWs grown by MBE  \cite{Jaffal2019, Harmand2013} and those grown by MOMBE \cite{Greenberg2014, Kelrich2015} are related to a different nucleation of steps, rather than to a different sticking of the adatoms to existing steps: an efficient nucleation in the case of MOMBE allows the formation of the cone-shaped NWs with a long tip terrace, while in MBE pencil-shaped NWs are obtained until the length (or a change of temperature) allows nucleation to take place.

\section{Conclusion}

A stable configuration is identified for the growth of a NW
with nucleation of steps at the base, by equating the  velocity of the
steps on the sidewalls, and the NW growth rate. The
quantities governing the shape are $R/a_\bot$ (radius at tip in
sidewall monolayer units), the desorption-limited diffusion length
$\lambda_{des}$, and a combination of parameters $\beta \tau /\lambda_{des}$ describing the overall efficiency of sticking at steps. In a microscopic approach, the two relevant parameters are the sticking probability / hopping probability ratio, $\frac{p_{inc}}{p_{hop}}$, and the desorption probability / hopping probability ratio, $\frac{p_{des}}{p_{hop}}$.

We identify two extreme domains: (1) if the sticking efficiency $\beta \tau /\lambda_{des}$, although non-zero, is below a threshold $\sim a_\bot/R$, the sidewall steps remain in the vicinity of the NW basis, and the axial growth is determined by
$\lambda_{des}$; this is the domain of cylinder-shaped NWs, and the conditions to choose in order to insert a quantum dot without the formation of parasitic dots on the sidewalls \cite{Harmand2013, Rueda2014, Wojnar2016} (2) for a sticking efficiency above this threshold, two segments coexist along the NW, an
array of equidistant steps (which govern the radial growth and the
cone angle) and a longer tip terrace (which governs the axial growth
rate). The NW shape then resembles that calculated for a specific case and described in Fig.~3~(d) of Ref.~\cite{Filimonov2016}. At very strong sticking, all parameters are determined by the sticking probability / hopping probability ratio. At moderate sticking, the two probability ratios contribute; when approaching the threshold $\beta \tau /\lambda_{des}\sim a_\bot/R$ from above, the length of the tip terrace  diverges
while the effective diffusion to the nanoparticle (or droplet) is
limited by desorption.

When applied to InP and ZnTe, the model points to a difference by an order of magnitude of the diffusion length of adatoms limited by desorption. This difference induces a different behavior of the axial and radial growth rates when adjusting the tapering angle.

Finally, from a practical point of view, the design of a waveguiding shell should take into account the presence of the tip terrace, since its length is of the same order as the wavelength for relevant values of the tapering angle (typically $2^\circ $). In the opposite limit, an upper bound is found for the tapering angle in the presence of a direct flux to the nanoparticle or droplet.

\section* {Acknowledgements}
The work at INL was done under the supervision of N. Chauvin,
    M. Gendry and P. Regreny. The Grenoble teams acknowledge funding by the French National Research
Agency (project Magwires ANR-11-BS10-013, COSMOS ANR-12-JS10-0002,
and ESPADON ANR-15-CE24-0029).

\appendix \section{Details of calculation}

\subsection{Establishing the equations}

We start from the well-known general solution of the BCF model. The adatom density $n$ on a wide terrace with an impinging flux $F_{\alpha}$ and a time to desorb $\tau$ is $n(x)=F_\alpha \tau$; on a terrace limited by two steps, $n(x)$ is usually looked for by subtracting a combination of two hyperbolic functions, $\cosh(x/\lambda_{des})$ and $\sinh(x/\lambda_{des})$, with prefactors determined by boundary conditions \cite{PimpinelliVillain, Pimpinelli1998, Peyla1998, Filimonov2014}.

Frequently assumed boundary conditions \cite{PimpinelliVillain} are either perfect sticking at the steps, or perfect sticking at one step and total reflectivity at the other step  (the so-called Ehrlich-Schw\"{o}bel effect). In the first case, $n=n_{eq}$ at both steps and we have only the cosh function with its extremum centered on the terrace; in the second case, $n=n_{eq}$ at one step and $\frac{dn}{dx}=0$ at the other step, and again there is only the cosh function, but centered on the reflecting step. Here we assume a perfect trap at the nanoparticle/droplet, and partial sticking at the steps \cite{Hata1991}: the regular terrace in the cone is symmetric, but the tip terrace is asymmetric, with different boundary conditions at both ends. The usual method for an asymmetric terrace is to add a $\sinh$ function \cite{Filimonov2014}, so that the adatom density is characterized by the length of the terrace and the two prefactors of the hyperbolic functions. However, any combination of the two hyperbolic functions can be written as a single, shifted hyperbolic function: if $a^2>b^2$, $a \cosh(x)+b \sinh(x)=\sqrt{a^2-b^2} \cosh(x+x_0)$, with $\tanh(x_0)=\frac{b}{a}$, and if $a^2<b^2$, $\sqrt{b^2-a^2}\sinh(x+x_0)$ with $\tanh(x_0)=\frac{a}{b}$. This allows us to look for the solution with a single hyperbolic function (here a cosh), with three parameters, the prefactor, and the distance of the extremum to each end of the terrace.

The adatom current on the terrace is given by $-D~\frac{dn}{dx}$. The incorporation into the nanoparticle/droplet must match this current. For incomplete sticking, the incorporation rate at a step is described \cite{Filimonov2014} by a current equal to $\beta~(n-n_{eq})$. Again, the incorporation at the step must match the current $-D~\frac{dn}{dx}$.

To sum up, the boundary conditions are:
\begin{itemize}
  \item at the NW tip, $(n-n_{eq})=0$,
  \item on each side of each step, $\beta~(n-n_{eq})=-D~\frac{dn}{dx}$.
  \end{itemize}

We look for $n(x)$ under the unique form (with a single cosh function),
\begin{equation}\label{EqA1}
n(x)=F_\alpha \tau-A (F_\alpha\tau-n_{eq})\frac{\cosh(x/\lambda_{des})}{\cosh(l_s/\lambda_{des})}.
\end{equation}

\begin{itemize}
  \item The terrace limited by two steps without Ehrlich-Schw\"{o}bel effect is a symmetrical system, hence we write Eq.~\ref{EqA1} with the origin at the terrace center. $A$ is determined by inserting this expression into the boundary condition at step stated above: the result is
\begin{equation}\label{EqA2}
A=\frac{\frac{\beta\tau}{\lambda_{des}}}{\frac{\beta\tau}{\lambda_{des}}+\tanh (\frac{l_s}{\lambda_{des}})}.
\end{equation}
This leads to Eq.~\ref{Eq1}. There is a current of adatoms $-D~\frac{dn}{dx}$ on each side of the step and the step velocity follows, Eq.~\ref{Eq2}.

\item For the tip terrace, using the same expression Eq.~\ref{EqA1} with the origin unknown, the boundary condition at the step, and the condition that this leading step has the same velocity as the other steps, determines the position of the origin at $l_s$ from the leading step, and the value of the prefactor $A$ (the same value as above, Eq.~\ref{EqA2}). The last parameter is the distance between the tip and the origin, $l_0$. The value of $l_0$ is obtained by inserting the expression for $n(x)$ into the boundary condition at tip, and the result is Eq.~\ref{Eq3}, which can be rewritten:
\begin{equation}\label{EqA3}
\cosh(\frac{l_0}{\lambda_{des}})=\cosh(\frac{l_s}{\lambda_{des}})\frac{\frac{\beta\tau}{\lambda_{des}}+\tanh(\frac{l_s}{\lambda_{des}})}{\frac{\beta\tau}{\lambda_{des}}}
\end{equation}
The contribution to the NW growth rate, Eq.~\ref{Eq4}, follows from the adatom current $-D~\frac{dn}{dx}$.
  \end{itemize}

\subsection{Using the equations}
The stability of the NW shape implies the equality between the NW growth rate $\frac{dL}{dt}$ and the step velocity $V_{step}$, with
\begin{eqnarray}\label{}
\frac{dL}{dt}&&=(F_\alpha-V_{sub}) \frac{2\lambda_{des} \tanh(\frac{l_0}{\lambda_{des}})}{R}+F~\kappa \label{EqA4}\\
V_{step}&&=(F_\alpha-V_{sub}) \frac{2\lambda_{des}}{a_\bot}\frac{
\beta\tau~
\tanh(\frac{l_s}{\lambda_{des}})}{\beta\tau+\lambda_{des}\tanh(\frac{l_s}{\lambda_{des}})}\label{EqA5}\\
&&=(F_\alpha-V_{sub})\frac{2\lambda_{des}}{a_\bot}\frac{\sinh(l_s/\lambda_{des})}{\cosh(l_0/\lambda_{des})}\label{EqA6}
\end{eqnarray}
The first equation is Eq.~\ref{Eq4}, the second one is Eq.~\ref{Eq2}, and the third one is Eq.~\ref{Eq2} modified using Eq.~\ref{Eq3}.

Eq.~\ref{EqA4} and Eq.~\ref{EqA6} contains the experimental data obtained on TEM images ($R$, $l_s$, $l_0$) and a single material parameter $\lambda_{des}$. The stability condition $\frac{dL}{dt}=V_{step}$, written with Eq.~\ref{EqA4} and Eq.~\ref{EqA6}, reads
\begin{eqnarray}\label{Eq7}
\frac{\sin \alpha}{\pi} \frac{2\lambda_{des}}{R}[\tanh(\frac{l_0}{\lambda_{des}})-\frac{R}{a_\perp} \sinh(\frac{l_s}{\lambda_{des}})\frac{1}{\cosh(\frac{l_0}{\lambda_{des}})}]\nonumber\\
+\kappa =0.
\end{eqnarray}
This is Eq.~\ref{Eq5}. A second order equation for $\tanh(\frac{l_0}{\lambda_{des}})$ is obtained by using the relation between $\cosh(\frac{l_0}{\lambda_{des}})$ and $\tanh(\frac{l_0}{\lambda_{des}})$. This equation can be used to derive $\lambda_{des}$ from experimental data; it can be used also to study the relationship between the tip terrace length and the cone angle, for a given value of $\lambda_{des}$, as done in section~\ref{compar}.

We can also calculate $l_s$ and $l_0$ as a function of the material parameters $\lambda_{des}$ and $\beta\tau$. This is particularly simple if $\kappa=0$. Then the stability condition using Eq.~\ref{EqA3} and Eq.~\ref{EqA5} is simply ${\sinh(l_0/\lambda_{des})=\frac{R}{a_\bot}}\sinh(l_s/\lambda_{des})$; combining that with $\cosh(l_0/\lambda_{des})$ from Eq.~\ref{EqA3}, we eliminate $l_0$ and obtain $\tanh(l_s/\lambda_{des})$, see the first line in Eq.~\ref{Eq6}. Then, $\tanh(l_0/\lambda_{des})$ follows, second line of (Eq.~\ref{Eq6}).

These results are used in Section~\ref{discussion}. The calculation with $\kappa\neq0$ gives also analytical results at the expense of a second order equation.



\begin{thebibliography} {5}

\bibitem{Claudon2010}
Claudon J, Bleuse J, Malik N S, Bazin M, Jaffrennou P, Gregersen N, Sauvan C,
Lalanne P and G\'{e}rard J.-M. 2010 \emph{Nat. Photonics} \textbf{4} 174

\bibitem{Reimer2012}
Reimer M E, Bulgarini G, Akopian N,
Hocevar M, Bavinck M B, Verheijen M A,
Bakkers E P A M,  Kouwenhoven L P and Zwiller V 2012 \emph{Nature Com} \textbf{3} 1

\bibitem{Dalacu2012}  Dalacu D, Mnaymneh K, Lapointe J, Wu X, Poole P J, Bulgarini G, Zwiller V and Reime M E 2012 \emph{Nano Lett.} \textbf{12} 5919

\bibitem{Haffouz2018}
Haffouz S, Zeuner K D, Dalacu D, Poole P J, Lapointe J, Poitras D, Mnaymneh K, Wu X,
Couillard M, Korkusinski M, Sch\"{o}ll E, J\"{o}ns K D, Zwiller
V and Williams R L 2018 \emph{Nano Lett.} \textbf{18} 3047

\bibitem{Zadeh2016} Zadeh I E, Elshaari A W, J\"{o}ns K D, Fognini A, Dalacu D, Poole P J, Reimer M E and Zwiller V 2016 \emph{Nano. Lett.} \textbf{16} 2289

\bibitem{Mnaymneh2019} Mnaymneh K, Dalacu D, McKee J, Lapointe J, Haffouz S, Weber J F, Northeast D B, Poole P J, Aers G C, Williams 2019 R L \emph{Adv. Quantum Tech} 1900021

\bibitem{Jaffal2019}  Jaffal A, Redjem W, Regreny P, Nguyen H-S, Cueff S, Letartre X, Patriarche G, Rousseau E, Cassabois G, Gendry M and Chauvin N 2019 \emph{Nanoscale} \textbf{11} 21847

\bibitem{Tchernycheva2007} Tchernycheva M, Travers L, Patriarche G, Glas F, Harmand J-C, Cirlin G E and Dubrovskii V G 2007 \emph{J. Appl. Phys.} \textbf{102} 094313

\bibitem{Dubrovskii2008} Dubrovskii V G, Sibirev N V, Cirlin G E, Tchernycheva M, Harmand J C and Ustinov V M 2008 \emph{Phys. Rev. E} \textbf{77} 031606

\bibitem{Dubrovskii2013} Dubrovskii V G, Timofeeva M A, Tchernycheva M and Bolshakova A D 2013 \emph{Semiconductors} \textbf{47} 50

\bibitem{Dubrovskii2006} Dubrovskii V G, Sibirev N V, Suris R A, Cirlin G E and Ustinov V M, Tchernysheva M and Harmand J-C 2006 \emph{Semiconductors} \textbf{40} 1075

\bibitem{Hofmann2008} Hofmann S, Sharma R, Wirth C T, Cervantes-Sodi F, Ducati C, Kasama T, Dunin-Borkowski R E, Drucker J, Bennett P and J Robertson 2008 \emph{Nature Materials} \textbf{7} 372

\bibitem{Wen2010} Wen C-Y, Tersoff J, Reuter M C, Stach E A and F M Ross 2010 \emph{Phys. Rev. Lett.} \textbf{105} 195502

\bibitem{Harmand2018} Harmand J-C, Patriarche G, Glas F, Panciera F, Florea I, Maurice J-L, Travers L and Ollivier Y 2018 \emph{Phys. Rev. Lett.} \textbf{121} 166101

\bibitem{Dongare2009} Dongare A M, Neurock M and Zhigilei L V 2009 \emph{Phys. Rev. B} \textbf{80}, 184106 

\bibitem{Wang2013} Wang H, Zepeda-Ruiz L A, Gilmer G H and Upmanyu M 2013 \emph{Nature Communications} \textbf{4}, 1956 

\bibitem{Checco2003} Checco A, Guenoun P and Daillant J 2003 \emph{Phys. Rev. Lett.} \textbf{91}, 186101 

\bibitem{Weijs2011} Weijs J H, Marchand A, Andreotti B, Lohse D and Snoeijer J H 2011 \emph{Phys. Fluids} \textbf{23}, 022001 

\bibitem{Schmidt2005} Schmidt V., Senz S and G\"{o}sele U 2005 \emph{Appl. Phys. A} \textbf{80}, 445 

\bibitem{Chen2001}  Chen X L, Lan Y C, Li J Y, Cao Y G and He M 2001 \emph{J. Crystal Growth} \textbf{222} 586

\bibitem{Plante2009} Plante M C and LaPierre R R 2009 \emph{J. Appl. Phys.}
\textbf{105} 114304

\bibitem{Filimonov2015} Filimonov S N and Hervieu Y Yu 2015 \emph{J. Cryst.
Growth} \textbf{427} 60

\bibitem{Filimonov2016} Filimonov S N and Hervieu Y Yu 2016 \emph{Russian Physics Journal} \textbf{59} 1206

\bibitem{Greenberg2014} Greenberg Y, Kelrich A, Calahorra Y, Cohen S and Ritter D 2014 \emph{J. Crystal Growth} \textbf{389} 103

\bibitem{Kelrich2015} Kelrich A, Dubrovskii V G, Calahorra Y, Cohen S and Ritter D 2015 \emph{Nanotechnology} \textbf{26} 085303

\bibitem{Harmand2013}
Harmand J-C, Jabeen F, Liu L, Patriarche G,
Gauthron K, Senellart P, Elvira D and Beveratos A 2013 \emph{J. Crystal Growth} \textbf{378} 519

\bibitem{Rueda2014} Rueda-Fonseca P, Bellet-Amalric E,
Vigliaturo R, Den Hertog M, Genuist Y, Andr\'{e}
R, Robin E, Artioli A, Stepanov P, Ferrand
D, Kheng K, Tatarenko S and Cibert J 2014 \emph{Nano Lett.}
\textbf{14} 1877

\bibitem{Rueda2016}
Rueda-Fonseca P, Orr\`{u} M, Bellet-Amalric E, Robin E, den Hertog
M, Genuist Y, Andr\'{e} R, Tatarenko S and Cibert J 2016
\emph{J. Appl. Phys.} \textbf{119} 164303

\bibitem{Chen2006} Chen C, Plante M C, Fradin C and LaPierre R R 2006 \emph{J. Materials Research} \textbf{21} 2801 	

\bibitem{BCF} Burton W K, Cabrera N and Frank F 1951 \emph{Phil. Trans. Roy. Soc.} \textbf{243}
299.

\bibitem{PimpinelliVillain} Pimpinelli A and Villain J 1998 Physics of
Crystal Growth (Cambridge University Press, Cambridge)

\bibitem{Hata1991} Hata M,
Isu T, Watanabe A, Kajikawa Y and Katayama Y 1991 \emph{J. Cryst. Growth} \textbf{114} 203

\bibitem{Pimpinelli1998} Pimpinelli A and Peyla  P 1998 \emph{J. Crystal
Growth} \textbf{183} 311

\bibitem{Peyla1998} Peyla P,
Pimpinelli A, Cibert J and Tatarenko S 1998 \emph{J. Crystal Growth} \textbf{184}
 75

\bibitem{Filimonov2014} Filimonov S N and Hervieu Y Yu 2014 \emph{e-J. Surf.
Sci. Nanotech.} \textbf{12} 68

\bibitem{Orru2018}
Orr\`{u} M , Robin E, Den Hertog M, Moratis K, Genuist Y,
Andr\'{e} R, Ferrand D, Cibert J and Bellet-Amalric E 2018 \emph{Phys. Rev.
Materials} \textbf{2} 043404

\bibitem{Glas2010} Glas F 2010 \emph{Phys. Statu Sol} B \textbf{247} 254

\bibitem{Dubrovskii2015} Dubrovskii V G,
Timofeeva M A, Kelrich A, Ritter D 2015 \emph{J. Cryst. Growth} \textbf{413} 25

\bibitem{Wojnar2012} Wojnar P,
Janik E, Baczewski L T, Kret S, Dynowska E, Wojciechowski T,
Suffczy\'{n}ski J, Papierska J, Kossacki P, Karczewski G,
Kossut J and Wojtowicz T 2012 \emph{Nano Lett.} \textbf{12} 3404

\bibitem{Harmand2010} Harmand J-C, Glas F and Patriarche G 2010 \emph{Phys. Rev.} \textbf{B 81} 235436

\bibitem{Wojnar2016} Wojnar P, P{\l}achta J, Zaleszczyk W, Kret S, Sanchez A M,  Rudniewski R, Raczkowska K, Szymura M, Karczewski G, Baczewski L T, Pietruczik A, Wojtowicz T and Kossut J 2016 \emph{Nanoscale} \textbf{8} 5720




\end{thebibliography}
\end{document}